# Running a Pre-Exascale, Geographically Distributed, Multi-Cloud Scientific Simulation

Igor Sfiligoi[1], Frank Wuerthwein[1], Benedikt Riedel[2] and David Schultz[2]

[1] University of California San Diego, La Jolla CA 92093, USA
[2] University of Wisconsin - Madison, Madison WI 53715 USA
`isfiligoi@sdsc.edu`

**Abstract.** As we approach the Exascale era, it is important to verify that the existing frameworks and tools will still work at that scale. Moreover, public Cloud computing has been emerging as a viable solution for both prototyping and urgent computing. Using the elasticity of the Cloud, we have thus put in place a pre-exascale HTCondor setup for running a scientific simulation in the Cloud, with the chosen application being IceCube's photon propagation simulation. I.e. this was not a purely demonstration run, but it was also used to produce valuable and much needed scientific results for the IceCube collaboration. In order to reach the desired scale, we aggregated GPU resources across 8 GPU models from many geographic regions across Amazon Web Services, Microsoft Azure, and the Google Cloud Platform. Using this setup, we reached a peak of over 51k GPUs corresponding to almost 380 PFLOP32s, for a total integrated compute of about 100k GPU hours. In this paper we provide the description of the setup, the problems that were discovered and overcome, as well as a short description of the actual science output of the exercise.

## 1 Introduction

In the past couple of years, there has been a lot of activity around getting ready for the exascale computing era. At the same time, public Cloud computing has been gaining traction, including funding agencies starting to invest in this sector; examples being NSF's ECAS and CloudBank awards, and the European Cloud Initiative. Cloud computing, with its promise of elasticity is the ideal platform for prototyping, as well as urgent computing needs. We thus attempted to demonstrate an exascale, or at least a pre-exascale workload using a real application that was already running on the Open Science Grid (OSG) [1], given that most workloads there follow the distributed High Throughput Computing (dHTC) paradigm, and could thus easily transition to the Cloud. The chosen application was IceCube's photon propagation simulation [2], for technical (heavy use of GPU at modest IO) and scientific reasons (high impact science). We emphasize that this meant it would not be a purely experimental setup, but it would produce valuable and much needed simulation for the IceCube collaboration's science program.

Since no public Cloud region, or even a single Cloud provider could deliver an on-demand exascale-class compute resource, we went for a geographically distributed multi-cloud setup, while still operating it as a single compute pool. Due to our limited budget, the compute exercise was executed as a short-lived burst, with the aim of demonstrating peak performance, even if only for a short amount of time. This allowed us to exceed 51k GPUs of various kinds in the pool at peak, which provided about 380 PFLOP32s (i.e. fp32 PFLOPS) and 160M GPU cores. We ramped to $2/3^{rd}$ of the total size in about half hour from the provisioning start and then reached the peak within about two hours. The total integrated compute time was of about 100k GPU hours.

The exercise used the IceCube's standard workload management system, i.e. HTCondor [3], but on dedicated hardware. We chose not to use the existing IceCube hardware installation since we did not want risk to disrupt the normal production activities, and we also wanted to minimize the risk of failure by using a slightly tuned setup. The used setup, including the special configurations and tunings are described in Section 2.

Section 3 provides an overview of the experience of ramping up to the peak 380 PFLOP32s and back down, including issues encountered in the process. We also provide an overview of the type of resources we were using at peak.

Section 4 describes the science behind the simulation application as well as a summary description of the simulation code internals. The effectiveness of the various GPU types for this specific application is also presented, both in terms of relative speed and total contribution during the run.

### 1.1 Related work

Running scientific workloads in the public Cloud is hardly a novel idea. This work is however novel in
   a) the concurrent use of all three major Cloud providers,
   b) the concurrent use of resources from all over the world,
   c) the concurrent use of several different GPU models,
   d) the total aggregate PFLOP32s at peak, and
   e) in its focus on the use of GPU-providing Cloud instances.

Moreover, running an unmodified, production scientific code in such a setup is also quite unusual.

In terms of sheer size, the largest scientific Cloud run we are aware of is the 2.1M vCPU weather modeling run performed by Clemson [4]. While the paper does not provide the achieved peak FLOP32s, it is unlikely it has exceeded 100 PFLOP32s, so it was significantly smaller than what we achieved in our setup. And while not exactly an apples-to-apples comparison, the 2.1M CPU cores are almost two order of magnitude fewer than the 160M GPU cores we provisioned at peak. Finally, it is also worth noting that their setup was confined to a single Cloud provider.

In terms of multi-Cloud setups, there have been many small scale, experimental studies. We are not aware of any other large-scale multi-Cloud scientific run though.

## 2 The workload management system setup

The workload management system software used for the pre-exascale, geographically distributed, multi-cloud setup was the same as normally used by IceCube to run its regular production on resources in the Open Science Grid and the European Grid Infrastructure, namely HTCondor. We saw no need to pick anything else, since HTCondor is naturally good at aggregating and managing heterogeneous resources, including when they are geographically distributed. HTCondor is also used for production activates at scales comparable to the desired peak, although mostly for CPU-focused workloads [5].

We decided to host a completely independent installation for the service processes, using dedicated hardware. We did this both to minimize the impact to IceCube's regular production environment, and to properly size it for the expected size and burst nature of the exercise, which are not typical of the abovementioned production environment. In the process we also tuned this setup to minimize the risk of failure during the actual multi-Cloud burst.

We also decided to not tackle the data movement problems to and from central storage related to such a large GPU compute burst. The IceCube jobs fetched input data from native Cloud storage and staged the results back to the native Cloud storage, too. In both cases we provisioned storage accounts close to where the jobs were actually running. For the input files we made educated guesses and used significant replication, since we did not know in advance where and how many GPU resources would be available. We fully acknowledge that this is not the ideal operations mode, but we were trying to tackle one problem at a time, namely large-scale, bursty GPU-heavy compute. We are likely to perform a more data-focused exercise in the near future.

Section 2.1 contains the summary overview of the HTCondor setup, including the reasons for the deployment choices. Section 2.2 provides the description of the changes needed to deal with Cloud native storage. Section 2.3 provides an overview of the problems encountered and the adopted solutions.

### 2.1 The multi-Cloud, geographically distributed HTCondor setup

IceCube, and the Open Science Grid community at large, has a lot of experience with operating large scale HTCondor instances. None however experiences the kind of burst growth we expected at peak times. The expected total resource pool was also comparable in scale to the largest production environments. Nevertheless, our experience suggested it was a feasible proposition, assuming we made arrangements to deal with a couple well-known bottlenecks.

The design goal for the HTCondor pool used in the multi-Cloud exercise was to be able to add at least 10k GPU resources per minute continuously up to a maximum of at least 120k concurrently running GPU jobs. In addition, most Cloud providers bundle many CPU cores with the GPU in their offering. We utilize this spare capacity by running at least a couple CPU jobs alongside each GPU job. This brings the total desired peak to about 350k concurrently running jobs.

HTCondor has been used in other production setups to sustain a job load of about 300k jobs, using CPU-only resources but with similar runtime characteristics. We applied the same sizing rules as they were applied there. The limiting factor in such setups is the number of running jobs that a single HTCondor job queue process, called the `schedd`, can handle; approximately 12k jobs. We thus used 10 nodes for GPU jobs and 20 nodes for CPU jobs. While the job queues are handled independently, they are scheduled as a single set and each of the jobs could still run on any compute resource that matched its requirements. By treating each `schedd` as a different logical user, the HTCondor policy engine, namely the `negotiator` process, keeps approximately the same number of jobs running on each of them by virtue of its standard fair-share policy.

The other well-known HTCondor limit is the ability of the central manager, namely the `collector` process, to add additional resources to its pool. The total size is actually not that important. The HTCondor central manager is essentially an in-memory database, and as long as there is enough RAM available, can easily scale to millions of resources being tracked. The problem is just the initial security handshake with the HTCondor daemon managing the remote resource, the so-called `startd`, and high-latency WAN communication makes the problem worse. It should also be noted that only the initial handshake is expensive. Further communication (in the absence of network problems) uses a much cheaper protocol.

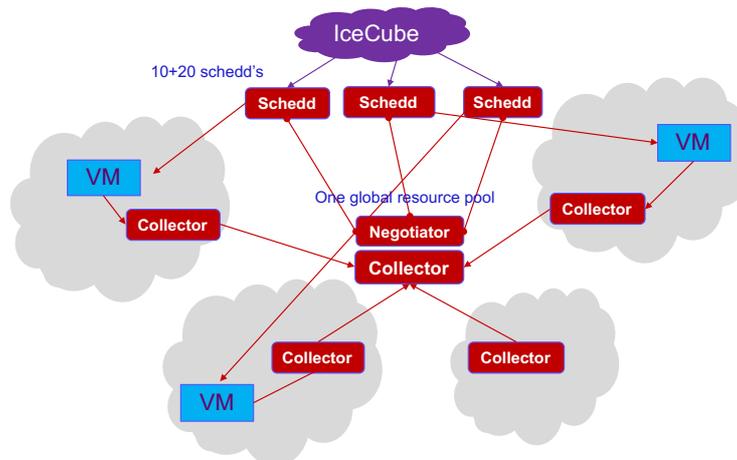

Fig 1 HTCondor setup with a tree of collectors.

Fortunately, HTCondor also provides a solution to work around the above limitation, by allowing for the creation of a tree of collectors, as seen in Figure 1. In this setup, the main `collector` process only ever talks to leaf `collector`s; it thus only has to establish a new security session at startup. Consequently, the `startd` processes are only aware of the leaf `collector`s and after picking one (at random), use it as their only connection to advertise their resources to the HTCondor pool. The leaf `collector`s then forward the information they receive to the main collector which thus gets a complete picture. The cost of resources joining the pool is thus distributed over many leaf nodes, and those leaf nodes can also be spread geographically to further minimize the cost inherent to high-latency WAN traffic. In our particular setup, we used one small Cloud instance running 20 leaf collector processes in each of the major Cloud regions, with `startd`s always picking one of those `collector`s in the closest region. The 20-per-node ratio was chosen as a perceived safe value to serve the largest of the Cloud regions and replicated everywhere in order to keep the setup as homogeneous as possible.

For completeness, the installed HTCondor version was 8.8.4, using the RPMs packaged and provided by OSG v3.4. We also used the standard OSG provided configuration unless otherwise specified. For securing the setup we used a shared secret, also known as a HTCondor pool password.

## 2.2 Dealing with data handling

Given both the unknown total size and composition of the compute pool and the bursty nature of the planed experiment, we decided to not add a data handling challenge of streaming data to and from IceCube's home storage to the exercise. We thus pre-staged all the needed input files into Cloud native storage and the outputs of the simulation were also to be staged back to native Cloud storage. In both cases, the native Cloud storage used was supposed to be close to where the job was running and belonging to the same Cloud provider that had provisioned the compute resource. We used extensive replication of input files to compensate for provisioning uncertainties. Fetching data back to IceCube's home storage happened asynchronously after the compute exercise was completed.

One problem of this approach is that the application would not know what the proper native Cloud storage to use is until it was running, including location and API used to access that storage. To address this, we had to solve two problems: 1) discover where the job is running and 2) make storage access homogenous for the application.

Discovering where a job runs is quite simple, once you know what Cloud provider you are using, and we used independent worker node images for each of the three providers, so the later was trivial. All three Cloud providers expose a Metadata service that allows for any process running in the instance to discover at runtime what region it belongs to. The syntax is slightly different between the three, but in all cases it is a REST API call, and is easy to automate e.g. by means of a *curl* invocation [6, 7, 8]. Once this information is known, a simple lookup table is all that is needed to find the appropriate storage URI.

In order to minimize the changes needed to the IceCube's compute jobs, we wrapped the metadata query, table lookup and the calls to the actual Cloud provider specific tools in a couple of simpler wrappers, one for downloading the data to the local disk and one for uploading the data from local disk to the closest native Cloud storage. The application thus only had to provide the relative paths, and the scripts would do the rest, making the application itself completely unaware of the environment it was running in. From the IceCube's jobs framework point of view, it was basically a two-line change of their production workflow on the Open Science Grid.

Finally, a quick mention on how we dealt with the actual application code binaries, which are indeed quite sizable. IceCube workflow expects their application software and dependencies to be present on the resources they land on. OSG provides IceCube with a uniform runtime environment via CVMFS, which uses just-in-time fetching of the binaries and extensive caching. This is efficient when the same application is run on a worker node multiple times. Our setup was however designed to be bursty, so the penalty of populating the cache may have been excessive. We thus decided to pre-loaded all the necessary software in the image itself. From an application point of view there was no difference at all.

We fully acknowledge that this is not the ideal data handling solution if one were to run this setup multiple times and over an extended period. We do plan to have a more data-oriented exercise in the near future.

### 2.3 Unexpected problems encountered in the HTCondor setup

Many of the compute resources IceCube usually runs on are behind NATs, and thus cannot be directly reached from the outside world. HTCondor has a solution for this problem, called Connection Brokering (CCB). It consists of a lightweight process that functions as a software-based router for `schedd`s to start remote compute jobs.

Public IPs are a metered commodity subject to quotas in at least some Cloud infrastructures. We were planning to use CCB for the Cloud setup, too. Unfortunately, during the initial feasibility tests, we discovered that HTCondor with CCB enabled does not properly handle rapid job startup bursts, resulting in an unusable setup, at least for our purposes. After convincing ourselves and the HTCondor development team that this was an actual software bug and not simply a configuration issue, and that a proper solution was not promptly available, we decided to just request public IPs for all the cloud instances that were part of the test pool. This turned out to be an acceptable solution for all the involved Cloud providers. It was more an annoyance than a real problem, but a limitation to be aware of for mimicking our setup.

We also encountered an issue in the policy engine code, resulting in jobs not being properly matched to resources. This bug is only triggered in very heterogeneous setups. Given our data placement policy, each Cloud region was treated as a different entity for scheduling purposes, as only a subset of the input data was placed in each Cloud region and the requirement to fetch from "local" Cloud native storage. Our pool was thus indeed quite inhomogeneous. Fortunately, the bug was part of a code optimization base that was introduced recently and could be avoided through a (mostly undocumented) configuration parameter:

```
NEGOTIATOR_PREFETCH_REQUESTS = False
```

## 3 The multi-Cloud, multi-Region setup

While the HTCondor workload management setup was almost identical to the production IceCube environment used on e.g. the Open Science Grid, the provisioning part of the exercise was completely ad-hoc. Nobody in our community had ever tried something similar before, so we did not have much prior art or engineering to rely upon.

On paper the problem did not seem too hard to tackle. All we needed to do was create an image containing the HTCondor binaries with proper configuration and start as many instances as we could in the shortest amount of time. The API is slightly different between the three Cloud providers, and the recommended methods and implementation limits slightly different, but apart from that the amount of work needed seemed to be very modest.

Given however the significant projected expense involved, and our relatively limited budget, we wanted to minimize the risk of hitting any technical limits during the actual GPU multi-Cloud burst. We thus spent significant time testing various aspects of the proposed setup using cheaper alternatives, namely CPU-only instances. We provide more details in section 3.4.

The tests did not identify any significant issues, but they did allow us to further tune our setup before the main exercise, which is described in section 3.2. It is worth noting that we did encounter unexpected problems during our actual run, which did not show up during testing. The decisive factor was likely the much higher variety of GPU Cloud instances compared to CPU Cloud instances.

We reached 65% of the maximum number of concurrently running instances in about half hour after we started the provisioning, 90% in about another half hour, and added an additional 10% over the course of the final hour. After about 2 hours from the provisioning start, we initiated a controlled shutdown which lasted another good hour. In section 3.3 we provide an overview of the resources provisioned at peak, as well as an analysis of the total compute integrated during the lifetime of the exercise.

The biggest hurdle we encountered during the setting up of the Cloud infrastructure for the bursty GPU run was not technology. Rather, the most time-consuming part was convincing the Cloud providers to even allow us to purchase that many GPU-enabled instances. A short summary is presented in section 3.1.

Finally, in section 3.5 we provide a short overview of the current Cloud pricing, with an emphasis on the resources we used during this run.

### 3.1 The social hurdle

Cloud computing is great, in that it allows anyone to provision resources with minimal effort and hardly any advance planning. This is however only true if your needs are small. Try provisioning any large amount of compute, and you will very fast bump into the limits imposed by the default quotas.

We can only speculate why this is the case. Our suspicion is that the Cloud providers want to minimize the risk that a user ends up with an unexpectedly large bill that they cannot pay. And as we demonstrate with our exercise, there is enough capacity out there to easily accumulate a million-dollar bill in just a couple of days!

Each Cloud provider has a different strategy on what an acceptable quota for different resources is. But in our case those quotas were universally too low. And our initial attempts at requesting sizable quota increases through standard channels were met by staff telling us that they did not have the authority to grant our requests.

We thus spent significant time reaching out to contacts at the three Cloud providers, explaining both the fact that we did have the money to pay the expected bill, were technically competent enough to not overspend and that our goal was actually worthwhile to support. We eventually got enough attention high enough in the leadership chain to get all our wishes granted, but it was a long process!

### 3.2 Provisioning the 51k GPUs over 3 Cloud providers using multiple regions

The main objective of this exercise was to provision the maximum number of GPUs that could be rented in any public Cloud for a couple of hours. We wanted to discover what was possible without any long-term commitments and demonstrate that it could be used efficiently for scientific workloads. The overlay workload management system, namely HTCondor, was described in the previous sections, so here we limit ourselves to the description of the provisioning part alone.

Given that HTCondor was dealing with job handling and resource matchmaking, the provisioning part was strictly limited to starting up properly configured worker node instances. In order to minimize external dependencies, we created self-contained and self-configuring images that we uploaded to the native Cloud image repositories. We decided to create a slightly different image for each of the three Cloud providers, both because there is no easy Cloud-native way of sharing images between them and it was easier to deal with system level differences between the three platforms. We did however use a single image for all of the regions belonging to each Cloud provider. The region-specific details were stored in a local lookup table. We then relied on their Metadata services for region discovery [6, 7, 8] and, after proper table lookups, finished the configuration of the node at instance boot time. All of this was done ahead of the run, has taken a couple of iteration to get right, but could be used as-is for any number of runs from that point on.

With images in the proper Cloud native repositories, it was just a matter of automating the rapid startup and teardown of a large number of instances. This meant not only dealing with three separate Cloud providers, but also specifics of each region. Any region is essentially its own island, with only the Identity and Access Management (IAM) tying them together. The automation thus had to deal with 28 independent Cloud regions, using 3 different APIs.

For Amazon Web Services (AWS), we were relying on the fleet mechanism [9]. We created one fleet configuration template for each GPU instance type in every targeted Cloud region, for a total of over 40 different templates. The provisioning in AWS was just a matter of creating new fleets with the desired number of instances. Note that the AWS API supports the request of multiple instance types within a fleet, but while that seemed to work fine for CPU instances in our preliminary tests, when mixing GPU instances it resulted in some instance types never being requested, so we opted for the safer although more complex alternative.

For Microsoft Azure, we used the Azure Virtual Machine Scale Sets [10]. Conceptually, an Azure Scale Set is very similar to an AWS Fleet. There are however a few significant differences; 1) an Azure Scale Set can be resized, AWS Fleets cannot, 2) an Azure Scale set has a hard upper limit to how many instances it can manage, while AWS Fleets do not, and 3) an Azure Scale Set always manages a uniform set of instances, while AWS Fleets can be more heterogeneous. Given that Azure Scale Sets can be easily resized yet have limited maximum size, we created several ahead of time for each GPU instance type in every targeted region, enough to allow us to provision all instances in even the most optimistic projection without coming too close to the hard limits of the Azure Scale Set infrastructure. Overall, we created over 200 of them. With the Scale Sets in place, the provisioning during the actual burst was as simple as increasing the instance count from 0 to the desired max.

In Google Cloud Platform (GCP) we used their Unmanaged Instance Groups [11]. They are conceptually very similar to the Azure Scale Set, so we used virtually the same approach; create a set of them ahead of time and then just set the instance number to the desired max during the exercise. GCP will then automatically deal with the rest.

The provisioning ramp-up went mostly smoothly. We did however encounter internal limits in a few Cloud regions that required manual intervention to recover. This can be observed in the two rapid rises in Figure 2a; the initial provisioning request at time zero, and the additional rapid rise after the manual recovery about 50 minutes later. Apart from that, the whole ramp-up happened completely autonomously. All Cloud regions across providers ramped up at the same time, as can be seen from the different colors in the figure, and no-one is ever providing a dominating fraction. Unfortunately, we are not authorized to identify the various regions involved. We can show which geographical areas were being provisioned, see Figure 2b. Here you can clearly see that instances have been joining the IceCube serving HTCondor pool from all over the world, with no single region significantly dominating the others.

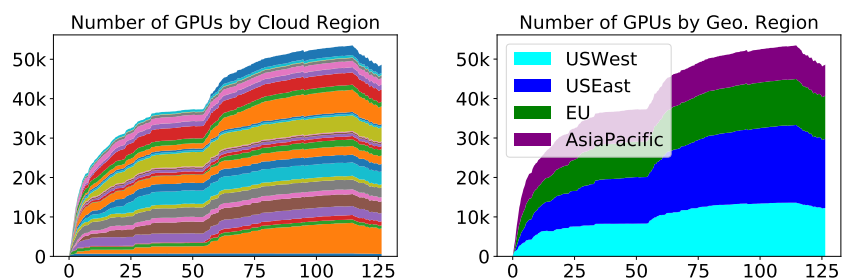

Fig 2 – Number of GPU instances over time (in mins) during the ramp-up period.
a) Grouped by Cloud region. b) Grouped by geographical region.

Managing a controlled shutdown turned out to be a harder problem than the provisioning. This was partially due to the desire of dynamically de-provision the instances only after the last job ran, and partially due to semantics of the provisioning tools used.

The desired sequence of the controlled shutdown was to remove all not-yet-started GPU jobs from the HTCondor queues, let the already running GPU jobs run to completion and de-provision the instance as soon as that GPU job terminated. Most instances also ran CPU jobs alongside the GPU jobs at all times. We would let CPU jobs run up until the last GPU job on the instance terminated and remove them at that point. Given the comparatively low value of the CPU-only compute, we wanted to optimize for maximum GPU utilization.

The removing of the not-yet-started GPU jobs from the HTCondor queues was obviously trivial. De-provisioning the running Cloud instances soon after the last GPU job ended was not. And we had to implement a different mechanism for each Cloud provider API.

The easiest was AWS. Since AWS Fleets were launched as ephemeral, all we had to do was to submit a set a system shutdown service jobs that would match as soon as the GPU job terminated. Unfortunately, we could not use this method for either Azure or GCP managed instances.

For Azure, the first obvious step was to set the desired numbers of instances in the Scale Set's configurations to the number of running instances at the beginning of the shutdown period, so no new instances would be started. However, instances in an Azure Scale Set have the property that just shutting down an instance at the system level will not de-provision that instance; one will still be charged for that time at the same rate as if the instance was still running [12]. The only proper way to de-provision an instance managed by an Azure Scale Set is to invoke the Azure API with the proper Scale Set and Instance ID pair. We thus wrote some service scripts that would extract that information from the instance Metadata service and automate the invocation of the proper API at the appropriate time.

For GCP, we had to follow a pattern similar to the one on Azure. First, we updated the desired number of instances for the Instance Groups, then we explicitly invoked the GCP API to de-provision each and every instance that we wanted removed from the Instance Groups. While the needed procedure is conceptually identical, the underlying reason is different. Unlike an Azure Scale Set, the semantics of an Instance Group is such that it will automatically replace a terminated instance with a new one, so explicit de-provisioning is needed to avoid an infinite start-and-terminate loop.

We had prepared and validated the above procedure during the initial testing phases and were quite confident we could manage the controlled shutdown with minimal effort. Unfortunately, we hit bugs in some Cloud regions, where our de-provisioning attempts resulted in Cloud APIs starting additional instances that could not be de-provisioned using the automated procedures we had in place. This resulted in a frantic troubleshooting session that ended by manually shutting down the offending instances manually through the Cloud provider's interactive Web portal. There were thousands of such instances, resulting in moderate waste that we did not anticipate. We are not authorized to identify neither the offending Cloud regions nor the associated Cloud provider(s). If we were to repeat such a large scale, bursty multi-Cloud exercise again, we would definitely put in place additional tools that would allow for rapid detection and cleanup of unexpected instances.

### 3.3 An overview of the provisioned resources

The total runtime of the exercise was slightly over 3 hours. We reached 90% of the maximum number of concurrently running instances in about 70 minutes after we started the provisioning. We sustained that level and added an additional 10% in the following 45 minutes. After which we started a controlled shutdown, which lasted just about an hour and a half. As can be seen from Figure 3a, we peaked at about 51.5k GPU jobs running.

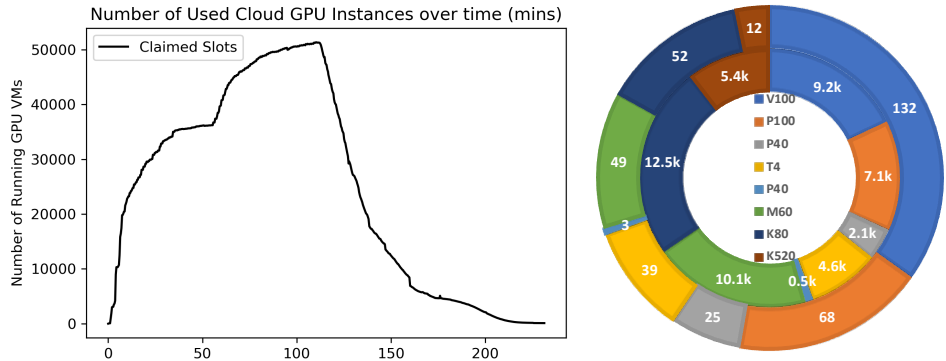

Fig 3 – a) Time evolution of the HTCondor pool. b) GPU composition at peak.
The inner circle shows the number of instances, the outer circle the PFLOP32s contribution.

The provisioned instances spanned 8 generations of NVIDIA GPUs. As seen from Table 1 and Figure 3b, the most abundant GPU types at peak were the K80 and the M60, but the most compute power was provided by the V100 and the P100 GPUs. We were also pleasantly surprised to see a significant contribution by the quite recent, and very cost effective T4.

**Table 1.** Distribution of GPU types at peak.

| GPU type (NVIDIA) | Count at peak | Total PFLOP32s |
|---|---|---|
| V100 | 9.2k (18%) | 132.2 (35%) |
| P100 | 7.1k (14%) | 68.1 (18%) |
| P40 | 2.1k (4%) | 25.2 (7%) |
| T4 | 4.6k (9%) | 38.6 (10%) |
| P4 | 0.5k (1%) | 2.5 (1%) |
| M60 | 10.1k (20%) | 48.8 (13%) |
| K80 | 12.5k (24%) | 51.6 (14%) |
| K520 | 5.4k (10%) | 12.4 (3%) |
| Total | 51.5k | 379.4 |

The total integrated time of the exercise was about 100k GPU hours. As seen from Table 2, it features the same NVIDIA GPU types, but is skewed toward the GPUs that ran the longest jobs, namely the M60s, since our controlled shutdown allowed any already started job to run to completion. Note that jobs on K80 and K520 GPUs were running on drastically smaller input files and were thus among the fastest to finish. Nevertheless, the M60 and the K80 were again the GPUs who contributed the most time, and the V100 and the P100 were the two GPUs who contributed the most compute power.

**Table 2.** Distribution of GPU types over the total runtime

| GPU type (NVIDIA) | Total walltime in hours | Total PFLOP32 hours |
|---|---|---|
| V100 | 18.2k (19%) | 260.7 (35%) |
| P100 | 16.0k (17%) | 152.8 (21%) |
| P40 | 4.4k (5%) | 51.5 (7%) |
| T4 | 7.3k (8%) | 61.4 (8%) |
| P4 | 0.7k (1%) | 3.6 (1%) |
| M60 | 25.1k (26%) | 121.3 (16%) |
| K80 | 18.5k (19%) | 76.2 (10%) |
| K520 | 7.0k (7%) | 16.1 (2%) |
| Total | 97.3k | 734.7 |

### 3.4 Preparations

Given the significant expected burn rate during the actual bursty multi-Cloud GPU exercise, we wanted to validate as many of our assumptions as possible using cheaper methods. Fortunately, the CPU-only instances are more than an order of magnitude cheaper and are a good proxy for how the infrastructure works. In the months leading to the actual exercise we tested three aspects of the whole system; 1) network performance, 2) HTCondor scalability and 3) Cloud API scalability.

The network part of the setup was deemed the most critical. If we were not able to transfer data in and out fast enough, we would have had huge waste in expensive GPU time, rendering the setup not feasible. We benchmarked both access to local cloud native storage as well as WAN transfers inside the Clouds and to scientific networks. The results were beyond our most optimistic expectations, with region-local Cloud storage easily exceeding 1Tbps and 100Gbps being the norm in the networking between Cloud regions. More than enough to make IO latencies negligible for our workload. Connectivity to scientific networks was also generally good, with 10 Gbps being exceeded in most setups. This made pre and post exercise data movement not at all challenging. The results of the tests have been presented at the CHEP19 conference[13].

The scalability of the HTCondor setup was technically the most critical one, but we had enough experience with other large-scale setups to be fairly confident that it would be feasible. Nevertheless, we did run a few bursty CPU-only tests to validate our assumptions, and did find some unexpected issues, as explained in Section 2.3. While unfortunate, these tests allowed us to implement workarounds that later tests proved to be effective under conditions expected during the actual GPU burst. At peak, our largest test setup had over 80k slots, coming from the same combination of Cloud providers and regions we were planning to provision from during the exercise.

13The major unknown was the performance of the Cloud APIs. We had virtually no prior experience with provisioning that many instances in any of the three Cloud providers, and we were also very eager to learn how fast these APIs could provision the resources. The preliminary tests did not show any problems, beyond the need of proper quotas being put in place, and the ramp-up speeds were very promising. Figuring out what the best APIs to use were was actually the major hurdle.

Finally, we also ran a small-scale GPU-enabled test, which tried to mimic as close as possible the final setup that would be used for the exercise. Here we discovered that AWS Fleets do not like heterogeneous GPU instance types, as mentioned in Section 3.2, and we adjusted accordingly.

### 3.5 Cloud cost analysis

Unfortunately, we are not authorized to discuss the actual price paid to carry out this exercise. We are instead providing an analysis based on the published list prices of the used Cloud providers, and in particular with a focus to opportunistic use scenarios. For clarity, the opportunistic use is called Spot pricing on AWS and Microsoft Azure, and preemptable instances in GCP. The rationale for using the opportunistic instance pricing stems from the fact that they are about three times cheaper that full price instances and dHTC workloads can gracefully recover from preemption, so there is no reason not to use them unless aiming for maximum resource pool size. Additionally, we only recorded an average preemption rate of 2% for opportunistic instances during this exercise, which is considered absolutely negligible in such setups.

Table 3 provides a list price range for instances providing each of the used GPU types. We also include an estimated hourly list cost for an experiment like ours, which turns out to be just shy of $20k/h. The table also provides a clear indication that it is much more cost effective to use the more recent GPU types; for example, the instances providing the K520 GPUs had approximately the same hourly list cost as instances providing T4 and P40 GPUs, yet they provided only a small fraction of the FLOP32s.

**Table 3.** Cloud opportunistic hourly pricing for various GPU types

| GPU type (NVIDIA) | Price range (list) | Count at peak | PFLOP32s at peak | Estimated list price at peak |
|---|---|---|---|---|
| V100 | $0.6-$1.0 | 9.2k | 132.2 | $7.2k/h |
| P100 | $0.4-$0.6 | 7.1k | 68.1 | $3.5k/h |
| P40 | $0.4-$0.6 | 2.1k | 25.2 | $1.0k/h |
| T4 | $0.2-$0.3 | 4.6k | 38.6 | $1.2k/h |
| P4 | $0.2-$0.2 | 0.5k | 2.5 | $0.1k/h |
| M60 | $0.2-$0.3 | 10.1k | 48.8 | $2.7k/h |
| K80 | $0.13-$0.3 | 12.5k | 51.6 | $2.9k/h |
| K520 | $0.2-$0.2 | 5.4k | 12.4 | $1.0k/h |
| Total | | 51.5k | 379.4 | $19.6k/h |

To be published in Proceeding of ISC High Performance 2020.

## 4 The IceCube science proposition

From the ground up, this exercise was planned with the goal of advancing science. While we were definitely interested in the technical aspect of the endeavor, we wanted primarily to demonstrate that such large-scale computation is actually useful for science. The technical participants in this group thus partnered early on with a scientific community that had a dHTC computing workload which needed significantly more compute resources than they were normally getting. This resulted in the exercise being a joint collaboration between personnel from both the Open Science Grid and the IceCube collaboration.

Section 4.1 provides an overview of the science behind IceCube. Section 4.2 provides an overview of the compute challenges involved in the IceCube science. Section 4.3 provides an overview of GPU-enabled code that was run during this exercise. Section 4.4 provides a comparison of the efficacy of the various GPU types for the purpose of IceCube, including how this skewed total contribution of various GPU types to the advancement of science objectives.

### 4.1 The IceCube Neutrino Observatory

The IceCube Neutrino Observatory [14] is the world's premier facility to detect neutrinos with energies above 1 TeV and an essential part of multi-messenger astrophysics. IceCube is composed of 5160 digital optical modules (DOMs) buried deep in glacial ice at the geographical south pole. Neutrinos that interact close to or inside of IceCube produce secondary particles, often a muon. Such secondary particles produces Cherenkov (blue as seen by humans) light as it travels through the highly transparent ice. Cherenkov photons detected by DOMs can be used to reconstruct the direction and energy of the parent neutrino. IceCube has three components: the main array, which has already been described; DeepCore: a dense and small sub-detector that extends sensitivity to ~10 GeV; and IceTop: a surface air shower array that studies O(PeV) cosmic rays.

IceCube is a remarkably versatile instrument addressing multiple disciplines, including astrophysics, the search for dark matter, cosmic rays, particle physics and geophysical sciences. IceCube operates continuously and routinely achieves over 99% up-time while simultaneously being sensitive to the whole sky. Highlights of IceCube scientific results is the discovery of an all sky astrophysical neutrino flux and the detection of a neutrino from Blazar TXS 0506+056 that triggered follow-up observations from a slew of other telescopes and observatories [15].

### 4.2 The importance of proper calibration

Any experiment requires a fundamental level of understanding of the employed instruments. In IceCube's case, the detector is built into a naturally existing medium, i.e. glacial ice, that has been deposited over millennia. *A priori* there was only limited information regarding the optical properties of the detector.

The optical properties of the glacial ice greatly affect the pointing resolution of IceCube. Improving the pointing resolution has two effects in this case: greater chance to detect astrophysical neutrinos and better information sent to the community. While IceCube can detect all flavors and interaction channels of neutrinos, about two-thirds of the flux reaching IceCube will generate a detection pattern with a large angular error, see Figure 4a. In the same figure you can also see that this angular error is mostly driven by systematic effects. Similarly, different optical models have a great effect on the reconstructed location of an event on the sky, see Figure 4b. The comparatively minute field of view of partner observatories and telescopes requires IceCube to provide as accurate as information as possible. Having the best calibration possible is therefore imperative.

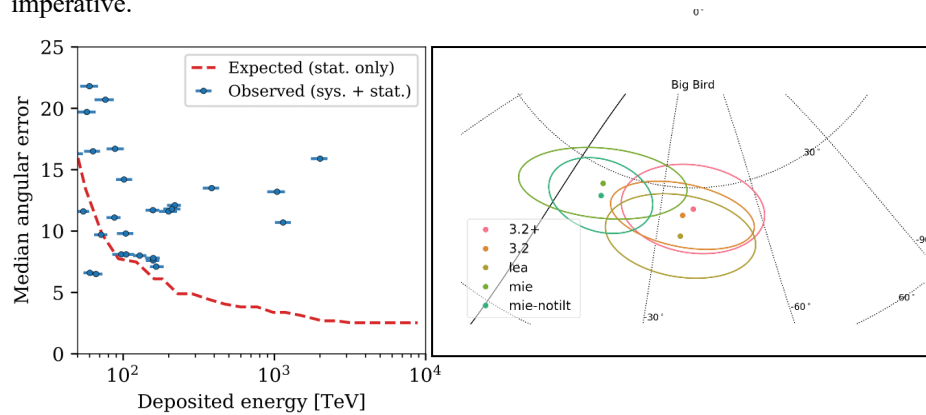

Fig 4 – Impact of the IceCube detector calibration on science results.
a) Angular momentum vs errors in IceCube b) Pointing area based on different estimates.

### 4.3  Using GPUs for photon propagation simulation

The photon propagation algorithm used by IceCube allows for massive parallelization using either a large number of CPU cores or GPUs [2]. The algorithm follows these steps. Initially a set of photons is created along the path of charged particles produced in the neutrino interaction or from *in-situ* light sources used for calibration. The number of photons inserted along the path depends on the energy loss pattern of the product. Most higher energy products will suffer stochastic energy loses due to bremsstrahlung, electron pair production, or ionization as they travel through the detector causing concentrations of light at certain points along the particle's path. For calibration sources, a fixed number of photons is inserted depending on the calibration source and its settings.

Once the location and properties of the photons have been determined, they are added to a queue. For a given device, a thread pool is created depending on the possible number of threads. If using a CPU, this typically is one thread per logical core. When using a GPU, this mapping is more complicated, but can be summarized as one to several threads per "core", and the exact mapping depends on the specific vendor and architecture. Each thread takes a photon out of the queue and propagates it. During the propagation, the algorithm will first determine the absorption length of the photon, i.e. how long the photon can travel before being absorbed. Then the algorithm will determine the distance to the next scatter. The photon is now propagated the distance of the next scatter. After the propagation, a check is performed to test whether the photon has reached its absorption length or intersected with an optical detector along its path. If the photon does not pass these checks, the photon is scattered, i.e. a scattering angle and a new scattering distance are determined, and the cycle repeats. Once the photon has either been absorbed or intersected with an optical detector, its propagation is halted and the thread will take a new photon from the queue.

The IceCube photon propagation code is distinct from others, e.g. Nvidia OptiX in that it is purpose-built. It handles the medium, i.e. glacial ice and the physical aspects of photon propagation in great detail. The photons will traverse through a medium with varying optical properties. The ice has been deposited over several hundreds of thousands of years. Earth's climate changed significantly during that time and imprinted a pattern on the ice as a function of depth. In addition to the depth-dependent optical properties the glacier has moved across the Antarctic continent and has undergone other unknown stresses. This has caused layers of constant ice properties, optically speaking, to be tilted and to have anisotropic optical properties.

### 4.4   The science output

The IceCube photon propagation code relies mostly on fp32 math, and this is why we focused on FLOP32s in the first three sections of this paper. There is however also a non-negligible amount of code that is needed to support the dataflow, and that cannot be directly tied to FLOPS of any kind. In the next few chapters we provide the actual measurements of how this translates in run times for the IceCube code.

**Table 4.** IceCube runtime for different GPU types.

| GPU type (NVIDIA) | Runtime in mins | Peak TFLOP32s | Correlation |
|:-:|:-:|:-:|:-:|
| V100 | 24 | 14 | 110% |
| P100 | 43 | 9.5 | 100% |
| P40 | 38 | 12 | 95% |
| T4 | 50 | 8.1 | 100% |
| P4 | 80 | 5.0 | 100% |
| M60 | 95 | 4.8 | 90% |
| K80 | 138 | 4.1 | 70% |
| K520 | 310 | 2.3 | 55% |
| GTX 1080 | 50 | 8.9 | 90% |

Before running the actual large-scale GPU burst, we benchmarked the IceCube photon propagation code, using a single representative input file, on all the various GPU-enabled instances of the various Cloud providers. Table 4 provides the observed runtimes, alongside the correlation to the nominal FLOP32s of those GPUs, relative to the performance of the recent and cost effective T4. A desktop-class GPU card, the NVIDIA GTX 1080 is also provided as a reference point. As can be seen, modern GPUs provide significantly shorter runtimes per nominal FLOP32 compared to the older ones.

The distribution of the observed run times of the jobs completed during the exercise matched very well the test data. The contribution to science results of the overall run was thus now much more skewed toward the newer GPU types, as can be seen from Figure 5. In particular, the older K80 and K520, which together contributed about a quarter of all the walltime, contributed less than 8% of the science output. In contrast, the more modern V100 and T4 together produced over 50% of all the science output while using about the same amount of wallclock time. A similar pattern can be also observed when comparing the list price versus the science output in the same figure.

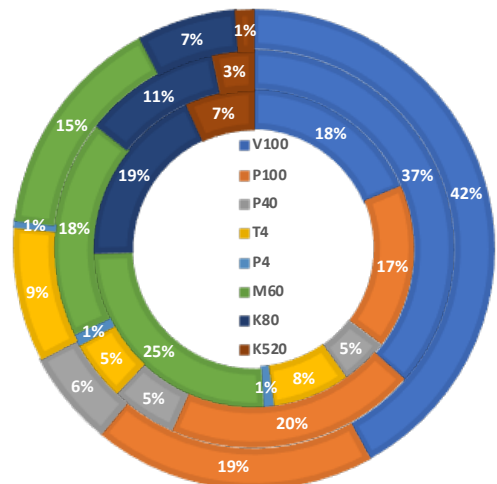

Fig 5 – Contribution of different GPU types during the exercise.
The external ring represents the fraction of science events simulated.
The intermediate ring represents the estimated list price fraction.
The inner ring represents the wallclock time fraction.

## 5 Conclusions

In this paper we present our experience in provisioning a pre-exascale dHTC workload using a real science application using Cloud resources from multiple Cloud providers. The chosen application was IceCube's photon propagation simulation, both for technical (heavy use of GPU at modest IO) and scientific reasons (high impact science). We emphasize that this meant it was not a purely experimental setup, but it produced valuable and much needed simulation for the IceCube collaboration's science program.

We managed to provision almost 380 PFLOP32s distributed over 51k instances using 8 different GPU types, from all over the world, reaching 2/3$^{rd}$ of the total size in about half hour from the provisioning start and the peak within about two hours. No special arrangements or long-term commitments were needed, apart from having the quotas raised to the appropriate levels. While the raising of quotas indeed involved a non-trivial effort, as explained in Section 3.1, we believe pre-exascale dHTC computing using Cloud resources is today within reach of anyone with a $100k budget.

Our exercise intentionally avoided to deal with real-time data movement problems in and out of the Cloud, by keeping all networking within Cloud providers' domains during the exercise. We do plan to have a follow-on exercise which will be much more data focused.

**Acknowledgements**

This work was partially funded by the US National Science Foundation (NSF) under grants OAC-1941481, MPS-1148698, OAC-1841530, OAC-1826967 and OAC-1904444.